\documentclass[letterpaper, twocolumn,superscriptaddress, showpacs,preprintnumbers,amsmath,amssymb] {revtex4}
\usepackage{graphicx}
\usepackage{dcolumn}
\usepackage{bm}
\usepackage{hyperref}
\hypersetup{
    colorlinks,
    citecolor=red,
    filecolor=blue,
    linkcolor=blue,
    urlcolor=blue
}

\begin{document}

\title{Ballistic spreading of entanglement in a diffusive nonintegrable system}

\author{Hyungwon Kim}

\affiliation{Physics Department, Princeton University, Princeton, NJ 08544, USA}

\author{David A. Huse}
\affiliation{Physics Department, Princeton University, Princeton, NJ 08544, USA}

\begin{abstract}
We study the time evolution of the entanglement entropy of a one-dimensional nonintegrable spin chain, starting from random nonentangled initial pure states.
We use exact diagonalization of a nonintegrable quantum Ising chain with transverse and longitudinal fields
to obtain the exact quantum dynamics.
We show that the entanglement entropy increases linearly with time before finite-size saturation begins,
demonstrating a ballistic spreading of the entanglement, while the energy transport in the same system is diffusive.
Thus we explicitly demonstrate that the spreading of entanglement is much faster than the energy diffusion in this nonintegrable system.
\end{abstract}


\maketitle

Entanglement is one of the unique features of quantum mechanics that does not exist in classical physics.
Originally quantum entanglement was viewed with some scepticism \cite{epr,es}, but recently the study of
entanglement has become a central part of many-body quantum physics and quantum information science.
Despite impressive recent progress in understanding entanglement from various viewpoints,
many of its aspects remain to be further explored.

One natural question about entanglement is its quantum dynamics under unitary time evolution.
If one starts an isolated quantum system in a nonentangled initial product pure state, how does the
entanglement grow with time?
Entanglement is not a conserved quantity like energy, that is transported.  Instead it is more like an infection or epidemic
\cite{omnes} that multiplies and spreads.  An initial state that is a product state has the information about the initial state of each
local degree of freedom (spins in our model below) initially localized on that degree of freedom.  Under the system's unitary
time evolution, quantum information about each spin's initial state can spread with time to other spins,
due to the spin-spin interactions.  This can make those spins that share this information entangled.

In real physical systems, information and entanglement can spread as fast as the speed of light (or sound).
For a lattice spin model, which lacks propagating light or sound,
an upper limit on the speed of any information
spread is given by the Lieb-Robinson bound,
which is set by the spin-spin interactions \cite{lieb} (for recent reviews, see Refs. \cite{ns,kliesch}).
For integrable one-dimensional models the entanglement does indeed spread ballistically
\cite{calabrese,chiara,maldacena}, which is to be expected since such systems
have ballistically propagating quasiparticles that can serve as carriers of the information.  For various localized models, on the other hand,
the entanglement has been shown to spread much more slowly, only logarithmically with time
\cite{chiara,znidaric,igloi,bardarson,vosk,spa,HO,osborne}.  In the present paper, we consider an intermediate case,
a quantum Ising spin chain that is neither integrable nor localized, whose energy transport is diffusive.

Here we investigate the spread of entanglement
in a diffusive nonintegrable system, at high temperature where
there are no ballistically propagating
quasiparticles and the only conserved quantity is the energy which moves diffusively \cite{integrable}.
Diagonalizing the entire Hamiltonian matrix, we numerically study the time evolution of the entanglement
and the diffusive dynamics of energy transport
for highly-excited thermal states of the system.
We show that the entanglement spreads ballistically,
while the energy moves only diffusively, and thus  slowly.
Although we choose a specific model Hamiltonian to study the quantum dynamics,
this result should be valid generally for nonlocalized and nonintegrable systems that do not have ballistically propagating quasiparticles or
long-wavelength propagating modes
such as light or acoustic sound.

As a simple nonintegrable model Hamiltonian, we choose a spin-1/2 Ising chain with both transverse and longitudinal fields.  The model is
translationally invariant, except at the ends of the chain, which we leave open.  Leaving the ends open allows the longest
distance within the chain to be its full length, so we can study energy transport over that distance, and the spread of bipartite
entanglement from the center of the chain to its ends.  If we had used periodic boundary conditions instead, the longest distances
that we could study would be cut in half.  Given the limited lengths that one can study with exact diagonalization, this factor of two
is quite important.
Our Hamiltonian is
\begin{align}
H = \sum_{i=1}^{L}g \sigma^{x}_{i} + \sum_{i=2}^{L-1}h\sigma^{z}_{i} +(h - J)(\sigma^z_1 + \sigma^z_L)+ \sum_{i=1}^{L-1} J\sigma_{i}^{z} \sigma_{i+1}^z ~.
\end{align}
$\sigma^{x}_i$ and $\sigma^{z}_i$ are the Pauli matrices of the spin at site $i$.
After searching a bit in the space of parameters to see where we have both fast entanglement spread and slow
energy diffusion and none of the terms singly dominates the energy spectrum,
we chose the longitudinal field $h = (\sqrt{5} + 1)/4 = 0.8090\ldots$ and the transverse field $g = (\sqrt{5} + 5)/8 = 0.9045\ldots$
and set the interaction $J = 1$
(and also set the Planck constant $\hbar$ to one); all results reported here are for these values.
Our qualitative results and conclusions do not depend on these parameter choices
as long as $g$, $h$ and $J$ are all of similar magnitude to each other to keep the system robustly nonintegrable \cite{parameters}.
Note that the magnitude of the energy ``cost'' to flip a spin in the bulk, from the applied longitudinal field and its interactions with its neighbors,
is $2h$ or $4J \pm 2h$. To keep the end sites similar in this respect to the bulk,
we reduce the strength of the longitudinal field on the end spins by $J$.
This is to avoid having some slow low-energy modes
near the ends that introduced small additional finite-size effects when we
applied the same magnitude of longitudinal field also to the end spins.

This Hamiltonian has one symmetry, namely inverting the chain about its center.  We always work with even $L$,
so the center of the chain is on the bond between sites $L/2$ and $(L/2)+1$.
This symmetry allows us to separate the system's state space into sectors that are even and odd under this parity symmetry,
and diagonalize within each sector separately.
Any mixed parity state can be obtained from a linear combination of even and odd parity states.
The statistics of energy level spacings within each parity sector of this nonintegrable Hamiltonian
should follow Gaussian orthogonal ensemble (GOE) statistics \cite{bohigas}. 
There are 32896 energy levels in the even sector for $L=16$, the largest system we have diagonalized.
Their level spacing statistics is in excellent agreement with the ``$r$ test'' introduced in Ref. \cite{vadim1} and the
Wigner-like surmise described in Ref. \cite{atas1}, as expected, indicating that this is indeed a robustly nonintegrable
model with no extra symmetries (see Supplement).

\begin{figure}
\includegraphics[width=3.375in]{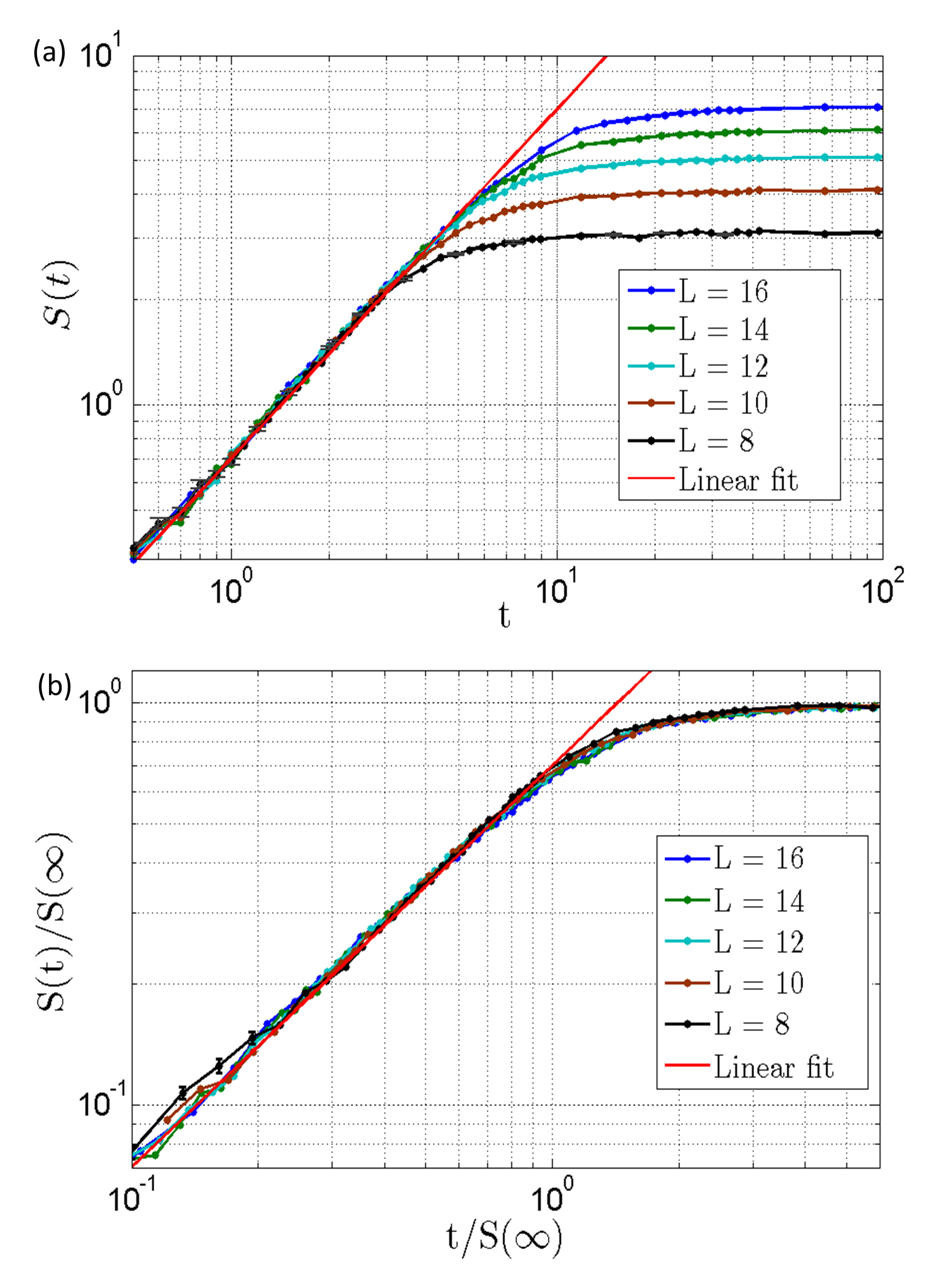}
\centering
\caption{(color online) (a) Spreading of entanglement entropy $S(t)$ for chains of length $L$.
Initially the entanglement grows linearly with time for
all cases, with the same speed $v \cong 0.70$.
Then the entanglement saturates at long time. This saturation begins earlier for smaller $L$, as expected.
The linear fit function is $f(t) = 0.70t$. Standard error is less than 0.04 for all points and thus the error bars are only visible at early times.
(b) Same data scaled by the infinite-time entropy for each $L$. Note that we use logarithmic scales both here and in Fig. 2.}
\label{ent}
\end{figure}

First, we consider the time evolution of the bipartite entanglement across the central bond between the two halves of the chain.
We quantify the entanglement entropy in bits using
the von Neumann entropy
$S(t) = -\rm{tr} \left[\rho_A(t) \log_2 \rho_A(t)\right] = -\rm{tr} \left[\rho_B(t) \log_2 \rho_B(t)\right] $
of the probability operators (as known as reduced density matrices) at time $t$ of either the left half ($A$) or the right half ($B$) of the chain.
As initial states, we consider random product states (with thus zero initial entanglement),
$|\psi(0)\rangle = |{\bf s}_1\rangle|{\bf s}_2\rangle...|{\bf s}_L\rangle$,
where each spin at site $i$ initially points in a random direction on its Bloch sphere,
\begin{align}
|{\bf s}_i\rangle = \cos\left(\frac{\theta_i}{2}\right)|\uparrow_i\rangle + e^{i\phi_i}\sin\left(\frac{\theta_i}{2}\right)|\downarrow_i\rangle ~,
\end{align}
where $\theta_i \in [0, \pi)$ and $\phi_i \in [0, 2\pi)$.
Such an initial state is in general neither even nor odd, and thus explores the entire Hilbert space of the pure states
as it evolves with unitary Hamiltonian dynamics.  This ensemble of initial states maximizes the thermodynamic entropy and thus corresponds to infinite temperature.
For each time $t$, we generate 200 random initial product states, let them evolve to time $t$, compute $S(t)$ for each initial state, and then average.
By doing so, the standard error at each time is uncorrelated.
The results are shown in Fig. \ref{ent}.
Ballistic linear growth of $S(t)$ at early time is clearly seen,
and the growth rate before the saturation is independent of $L$.
[Note, there is an even earlier time regime at $t\ll 1$ where the entanglement initially grows as $\sim t^2|\log{t}|$;
this regime is just the initial development of some entanglement between the two spins immediately adjacent to the central bond.]

In the long time limit, the time evolved state, on average,
should behave like a random pure state (a random linear combination of product states) \cite{entropy}.
In Ref. \cite{page}, it is shown that the average of the entanglement entropy of random pure states is
\begin{align}\label{ent_limit}
S^R = \log_2 m - \frac{m}{2n\ln 2} - {\cal O}\left(\frac{1}{mn}\right)~.
\end{align}
where $m$ and $n$ are the dimension of the Hilbert space in each subsystem, with $m \leq n$.
Since $m = n = 2^{L/2}$ in our case, $S^R \simeq \frac{L}{2}$ in the large $L$ limit.
This limiting value indicates that
the entanglement spreads over the entire subsystem of length $L/2$.
Therefore, before saturation begins, we can interpret $S(t)$ (in bits) as
a measure of the distance over which entanglement has spread, and
its growth rate thus as the speed of the ballistic entanglement spreading.
It is clear from figure \ref{ent}(a) that at long time ($t > 20 \sim 100$ depending on the system size) $S(t)$ saturates close to $S^R$.
We found that the deviation of the saturation value from $L/2 - 1/(2\ln 2)$ (Eq. \ref{ent_limit})
is small ($\sim$ 0.19 for $L$ = 8 and $\sim$ 0.11 for $L$ = 16).
Since the entanglement entropy saturates because of the finite length $L$,
this deviation from $S^R$ is a correction to the leading finite-size effect, which should be negligible in the thermodynamic limit.

This behavior suggests the finite-size and finite-time scaling form for the entanglement entropy:
\begin{align}\label{fsS}
S(t) \approx S_L(\infty) F(t/S_L(\infty)) ~,
\end{align}
where $S_L(\infty)$ is the infinite-time average value of the entanglement entropy \cite{longtime_ent} for chain length $L$,
the scaling function $F(x)\sim vx$ for $x\rightarrow 0$ ($v$ is the spreading rate), and $F(x)\rightarrow 1$ for $x\rightarrow\infty$.
Figure \ref{ent}(b) confirms that this scaling works well.

\begin{figure}
\includegraphics[width=3.375in]{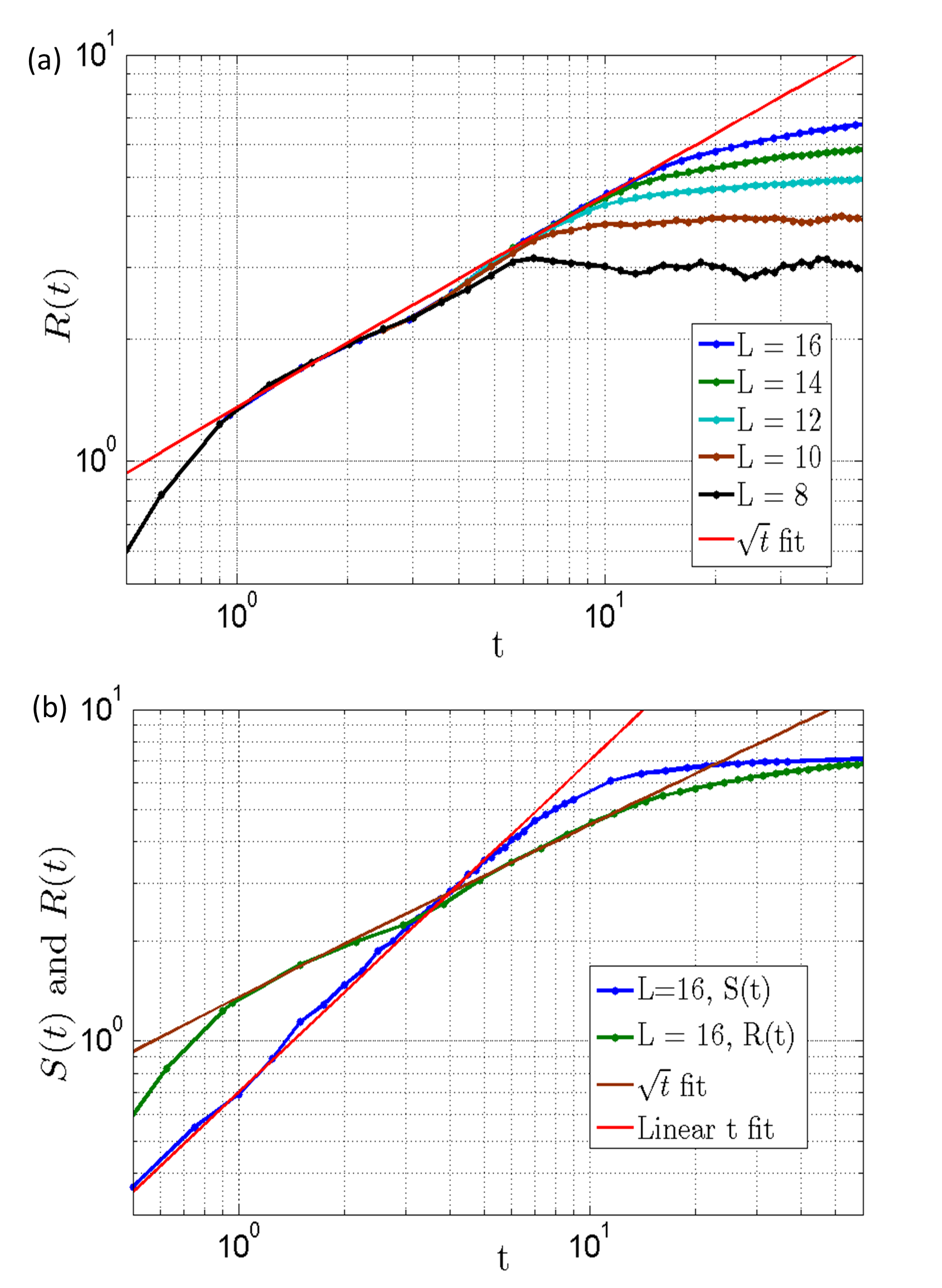}
\centering
\caption{(color online) (a) The average energy spreading, $R(t)$ (defined in the main text) vs. time. Before saturation, its behavior does not depend on the system size. As we increase the system size, diffusive $\sqrt{t}$ behavior becomes more apparent. (b) Direct comparison of $S(t)$ and $R(t)$ for $L=16$. It is clear that the entanglement spreads faster than energy diffuses in the scaling regime before saturation.}
\label{diff}
\end{figure}

Now let's consider the diffusive dynamics of this system.
As an example, we study the diffusive spreading of an initially localized energy inhomogeneity.
First we prepare the system in the maximal thermodynamic entropy mixed state (equilibrium at infinite temperature)
and put a small energy perturbation on the center bond.
Then we observe how this extra local energy spreads over the system under unitary time evolution.
Specifically, the initial probability operator (density matrix) is
\begin{align}
\rho(0) = \frac{1}{2^L}\left(I + \epsilon \sigma^z_{L/2}\sigma^z_{L/2 + 1}\right)~,
\end{align}
where $I$ is the identity operator and $\epsilon$ is a small number.  Note that $I$
commutes with $H$, so we only need to time evolve the perturbation.
Then, we compute the local energy, $\langle H_r\rangle(t)$, at each site and bond $r$ at time $t$.
The position index $r$ is an integer (1 to L) for each site and a half-integer (3/2 to L - 1/2) for each bond.
Explicitly,
\begin{align}
H_r = \left\{
\begin{array}{l l}
g\sigma^x_r + h\sigma^z_r &\quad \text{sites  $2\leq r \leq L-1$} \\
g\sigma^x_r + (h-J)\sigma^z_r & \quad \text{$r = 1$ or $L$} \\
J\sigma^z_{r-1/2}\sigma^z_{r+1/2} & \quad\text{bonds $3/2\leq r \leq L-1/2$} ~.
\end{array}
\right.
\end{align}
This is just a decomposition of the hamiltonian $H = \sum_r H_r$.
Trivially, $\langle H_{r}\rangle = \epsilon \delta_{r, \frac{L+1}{2}}$ at time $t=0$.
To quantify the energy spreading at time $t$, we compute an average ``distance'' $R(t)$ that the energy has moved away from the center bond:
\begin{align}\label{R}
R(t) = \frac{2}{\langle H \rangle}\sum_r \left|r - \frac{L+1}{2}\right| \langle H_r\rangle(t) ~,
\end{align}
where $\langle H \rangle=\epsilon$ is the conserved total energy.
Figure \ref{diff}(a) is the plot of $R(t)$ for L = 8, 10, 12, 14, and 16.
If the extra energy at long time is distributed equally to all sites and bonds,
$R(\infty) \rightarrow \frac{L}{2}\frac{2L-2}{2L-1} \simeq \frac{L}{2}$ and thus close to the maximum value of entanglement spreading
(the factor of 2 in Eq. \ref{R} is to make the long time value of $R(t)$ comparable to that of $S(t)$).
We find that the saturation value, $R(\infty)$ \cite{longtime_diff},
grows linearly with the system size but is always slightly smaller than $L/2$ due to the final local energy distribution not being
homogeneous near the ends of the chain.

If this dynamics is diffusive, the energy spread is $R(t) \approx \frac{4}{\sqrt{\pi}}\sqrt{Dt} \sim \sqrt{t}$ (one-dimensional random walk)
for sufficiently large $t$ ($t \geq 1 $ in our case) before finite-size saturation begins.
$D$ is the energy diffusivity, which only depends on the interaction parameters, not the system size.
Figure \ref{diff}(a) clearly shows that $R(t)$ is independent of system size at early stages,
and it grows as $\sim\sqrt{t}$
before saturation begins.
For L = 8, the frequency scale of the many-body level-spacing is
of order 0.1
and thus $R(t)$ begins oscillating around $t \sim 10$.
Although the system sizes that we can diagonalize are not large enough to
show a wide range of time scales,
they do show that the speed of entanglement spreading becomes faster than the rate of diffusive energy spreading
by direct comparison of $S(t)$ and $R(t)$.
Figure \ref{diff}(b) is the plot of $S(t)$ and $R(t)$ for L = 16.
In the very beginning ($t\leq 1$), $R(t)$ grows faster than $S(t)$ due to
microscopic details of the dynamics, but soon the linearly-growing $S(t)$ overtakes $R(t)$
and approaches its saturation while $R(t)$ is growing only as $\sim\sqrt{t}$.
Therefore, this is a direct demonstration of the contrast between ballistic entanglement spreading and diffusive energy transport.

In conclusion, we have demonstrated
that quantum entanglement spreads ballistically in a nonintegrable diffusive system.
Since there are no ballistically traveling quasiparticles, the mechanism of entanglement spreading is
different from what happens in integrable systems, where these quasiparticles can carry both energy and information.
At high enough temperature,
almost all states are relevant to the dynamics,
and the dynamics is constrained by only few conservation laws (in our case, only the total energy).
In this regime, the concept of quasiparticles is not well-defined for the system we have studied.
Even so, if we do heuristically describe the dynamics of our diffusive model in terms of quasiparticles,
these quasiparticles scatter strongly and frequently and thus have a short mean free path.
This limits the energy transport to be diffusive.
But apparently the quantum information needed to spread
entanglement is passed along in each collision, presumably to all outgoing quasiparticles from each collision.
Thus this information spreads
in a cascade or shower of collisions and the edges of this shower spread ballistically.

We conjecture that for highly-excited nonintegrable systems such as we study here, there are no \emph{local} observables whose
correlations spread more rapidly than diffusively, even though the entanglement spreads ballistically.
Note that this is a strong conjecture that goes well beyond what we can test numerically.

We have used the analogy from Ref. \cite{omnes} between the spreading of entanglement and the spreading of an
epidemic.  But it is an unusual sort of nonlocal epidemic, where the symptoms of the ``disease''
can not be detected by any local observables.  In Ref. \cite{liu} they make an analogy instead to a tsunami;
again this appears to be a very gentle nonlocal tsunami, whose effects can only be detected by nonlocal observables.
An interesting question
that we leave for future work is:  What is the
simplest and most local operator that can detect this ballistically spreading entanglement?  We detected it using the state of the full
system, but if the entanglement has only traveled a distance $\ell$ in each direction from the central bond, it should be detectable
by some operators that only involve the spins within that distance.

We thank Joel Moore and Michael Kolodrubetz for discussions. We also acknowledge Hanjun Kim, Seok Hyeong Lee, Taewook Oh, and Jonathan Sievers for fruitful discussions about numerics. The simulations presented in this work were performed on computational resources supported by the Princeton Institute for Computational Science and Engineering (PICSciE) at Princeton University. H. K. is partially supported by Samsung Scholarship.  This work was supported in part by NSF under DMR-0819860 and by funds from the DARPA Optical Lattice Emulator program.

\section{Sumpplementary Material}

\subsection{Choice of Hamiltonian parameters}
\begin{figure}
\includegraphics[width=3.375in]{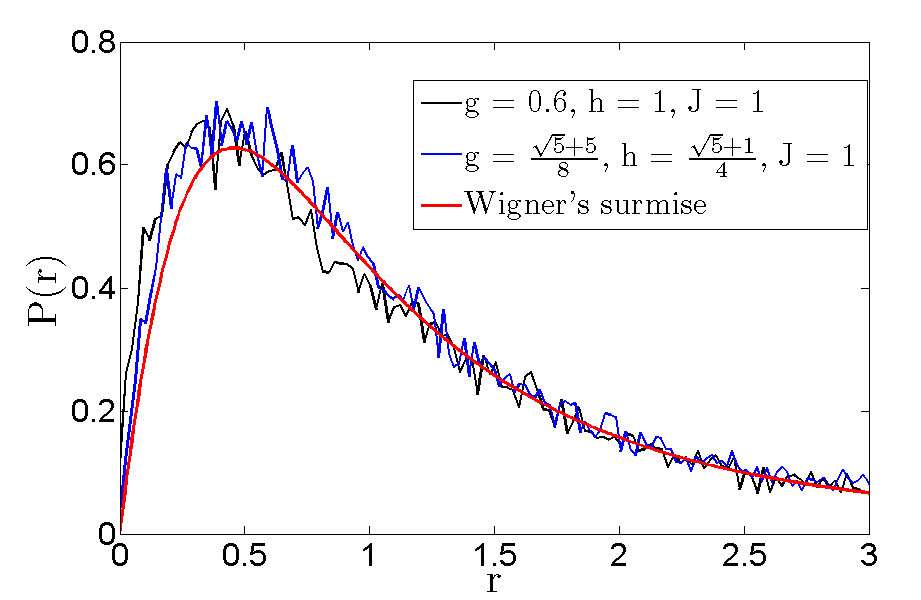}
\centering
\caption{(color online) Ratio distribution of level spacing for two sets of parameter choices. $r$ is the ratio of level spacings between two adjacent energy gaps.
For $L$ = 16, there are 32896 eigenvalues in even parity eigenstates, from which we obtain 32894 ratios. This ratio distribution is computed from the histogram of the lowest 32000 ratios with 250 equally spaced bins.
$(g, h, J) = ((\sqrt{5} + 5)/8, (\sqrt{5} + 1)/4, 1.0)$ case shows better level repulsion near $r = 0$. }
\label{r_dist}
\end{figure}

The model we consider is the one dimensional Ising chain with transverse and longitudinal field with open boundary condition.
\begin{align}\label{ham}
H = \sum_{i=1}^{L}g \sigma^{x}_{i} + \sum_{i=2}^{L-1}h\sigma^{z}_{i} +(h - J)(\sigma^z_1 + \sigma^z_L)+ \sum_{i=1}^{L-1} J\sigma_{i}^{z} \sigma_{i+1}^z ~.
\end{align}
Strictly speaking, this model is nonintegrable as long as all parameters, $g$, $h$, and $J$, are nonzero.
However, since the accessible size of exact diagonalization method is limited, some choices of parameters can be
superior than other choices to do finite size scaling.
One characteristic feature of a nonintegrable system is that its eigenvalues show level repulsion \cite{mehta_supp}.
One convenient quantity to look at its level statistics is the ratio distribution of adjacent eigenvalues \cite{vadim_supp, atas_supp}.
First, order the eigenvalues in ascending order, then compute the distribution of $r = (\lambda_{i+2} - \lambda_{i+1})/(\lambda_{i+1} - \lambda_{i})$,
where $\lambda_i$ is the $i$'th eigenvalue.
For a nonintegrable system, this distribution should follow the Wigner-like surmise while an integrable system
should exhibit Poissonian distribution \cite{atas_supp}.
Figure \ref{r_dist} is the ratio distributions for two different set of parameters, $(g, h, J) = (0.6, 1.0, 1.0)$ and $(g, h, J) = ((\sqrt{5} + 5)/8, (\sqrt{5} + 1)/4, 1.0)$, within even sector of $L = 16$. Although both of them are far from Poissonian and look similar to Wigner's surmise,
$(g, h, J) = ((\sqrt{5} + 5)/8, (\sqrt{5} + 1)/4, 1.0)$ case shows stronger level repulsion and agrees with the Wigner's surmise better.
To minimize finite size effect, the latter choice of parameter should be more appropriate.

\begin{figure}
\includegraphics[width=3.375in]{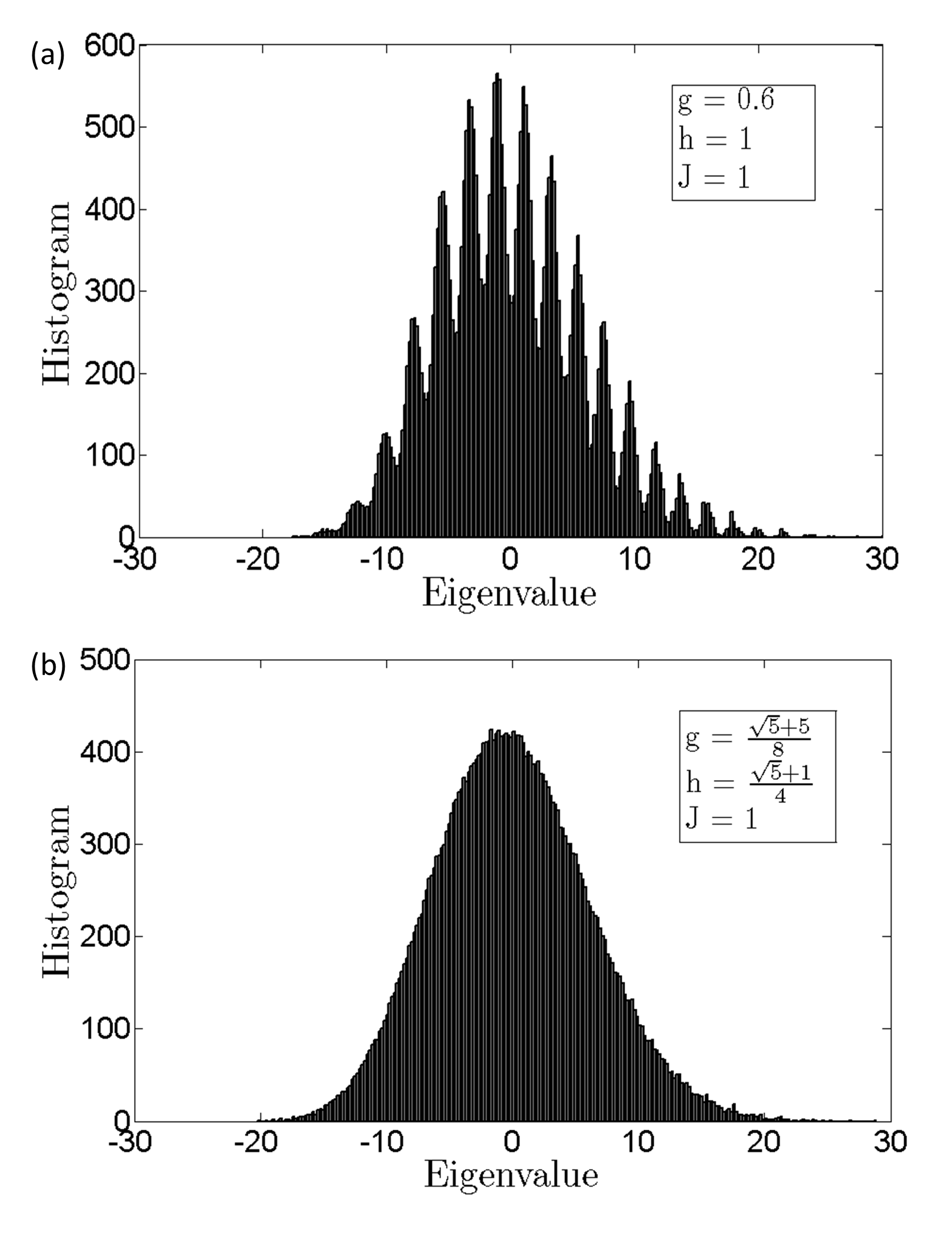}
\centering
\caption{Histogram of 32896 even state eigenvalues for two sets of parameters. $(g, h, J) = (0.6, 1.0, 1.0)$ case has quasi-periodic structure whereas the other case is featureless. In order to study generic properties of nonintegrable systems with limited accessible system size, we should use parameters with which the distribution of eigenvalues does not have distinct structure.}
\label{ev_hist}
\end{figure}

Another aspect to consider is the structure of eigenvalues. For a generic nonintegrable model, we expect the distribution of eigenvalues to be featureless.
Figure \ref{ev_hist} is the histogram of 32896 even state eigenvalues for the same two sets of parameters with $L = 16$.
Clearly, $(g, h, J) = (0.6, 1.0, 1.0)$ case shows the reminiscent feature of $g = 0$, quasi-periodic eigenvalues
with small perturbation coming from nonzero $g$,
which could make the finite size scaling very difficult under the limited accessible system size.
Therefore, $(g, h, J) = ((\sqrt{5} + 5)/8, (\sqrt{5} + 1)/4, 1.0)$ is a better choice of parameters.
Note that checking only one of the criteria used here might not be sufficient to accept the parameters,
as Figure \ref{r_dist} shows that $(g, h, J) = (0.6, 1.0, 1.0)$ option may not appear too bad
and the distribution of eigenvalues of the integrable case ($h = 0$) seems to be very similar to that of
$(g, h, J) = ((\sqrt{5} + 5)/8, (\sqrt{5} + 1)/4, 1.0)$.

\begin{figure}
\includegraphics[width=3.375in]{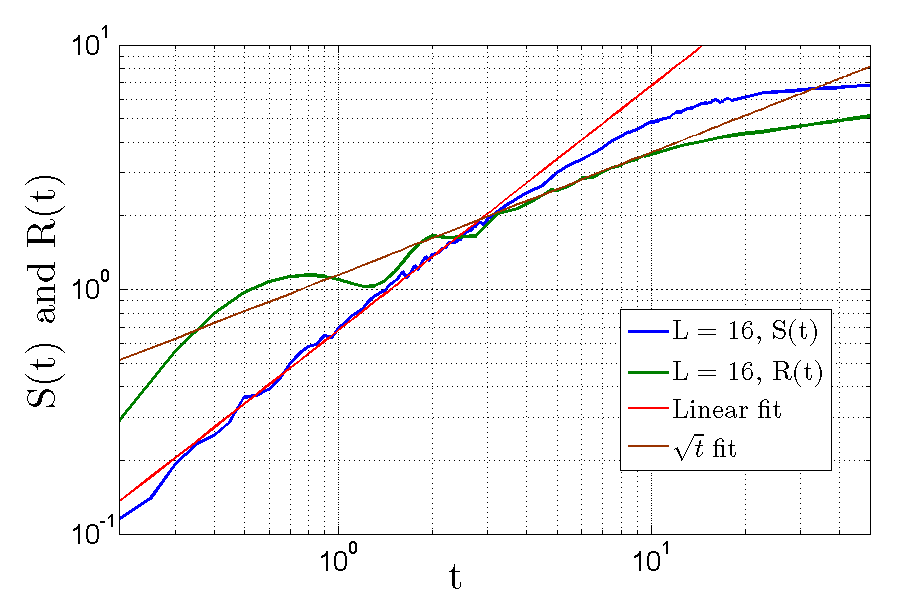}
\centering
\caption{(color online) Direct comparison of the entanglement spreading $S(t)$ and the energy diffusion $R(t)$ for $L = 16$ and
$(g, h, J) = (-1.45, \pi/2, 1.0)$. Entanglement spreading is faster than energy diffusion but direct comparison is more difficult than Figure \ref{diff} (b) in the main text. }
\label{half_pi}
\end{figure}

\begin{figure}
\includegraphics[width=3.375in]{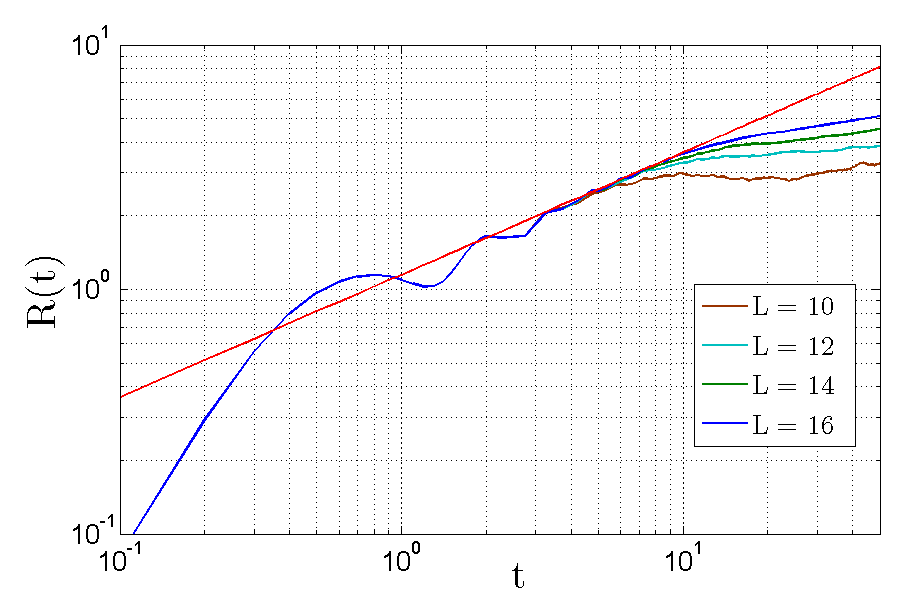}
\centering
\caption{(color online) Energy diffusion measure $R(t)$ (defined in the main text) for $(g, h, J) = (-1.45, \pi/2, 1.0)$.
The initial transient does not scale with the system size and thus it is irrelevant in thermodynamic limit. Before finite size effect begins to matter,
it shows diffusive behavior which extends as the system size increases.}
\label{diff_half_pi}
\end{figure}

As long as the magnitude of $g$ and $h$ are comparable, most, if not all, choices of parameters satisfy the above two criteria.
Here we present the result obtained from one choice of them; $(g, h, J) = (-1.45, \pi/2, 1.0)$.
Figure \ref{half_pi} is the central plot of the main text using this set of parameters.
It shows the same qualitative features; ballistically spreading entanglement, initial wiggle of energy diffusion, diffusive transport of energy,
and the same final saturation value of entanglement.
However, compared to the choice given in the main text,
this option yields earlier deviation from ballistic spreading of entanglement
and especially extended initial transient in energy diffusion.
What mostly limits the choice of parameters is the irregular initial behavior of energy diffusion
while spreading of entanglement has appeared relatively insensitive to the choice of parameters.
Figure \ref{diff_half_pi} shows that the initial non-diffusive behavior
does not scale with the system size and thus does not affect any asymptotic properties.
Although it is still apparent that the entanglement spreading is faster than the energy diffusion,
the directly comparable range is smaller than that in the main text.
This is the main reason of the parameter choice in the main text.

We emphasize that the reason for the careful choice of parameters is mainly due to
the restricted accessible system size of exact diagonalization method.
We can still see the ballistic spreading of entanglement
and diffusive behavior of energy transport even with the parameter choice of $(g, h, J) = (0.6, 1.0, 1.0)$.
This set of parameters just gives small scaling range and very strong initial transient in energy diffusion
that makes direct comparison difficult.

\subsection{Saturation value of long time entanglement}

\begin{figure}
\includegraphics[width=3.375in]{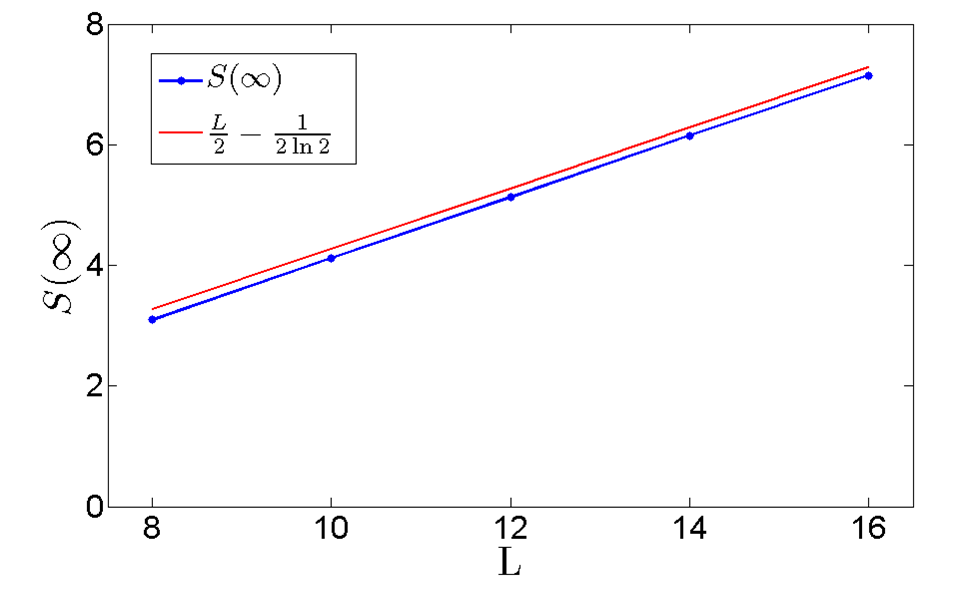}
\centering
\caption{(color online) Comparison of long time values from numerical data with expected values. The difference is very small and entanglement will grow indefinitely in thermodynamic limit. The parameters of Hamiltonian is same as those in the main text, $(g, h, J) = ((\sqrt{5} + 5)/8, (\sqrt{5} + 1)/4, 1.0)$. One standard error is of order 0.02 thus omitted.}
\label{end_ent}
\end{figure}

We present the explicit comparison between the average entanglement entropy of random pure states $S^R = L/2 - 1/(2\ln2)$, derived in Ref.\cite{page_supp}
and saturation value $S(\infty)$. Figure \ref{end_ent} clearly shows that they are very close to each other and grows linearly with the system size.
In a large size limit, the correction to the leading part $L/2$ should become negligible.
Since the change of difference, $S^R - S(\infty)$, from different system size $L$ is not well separated outside the standard error,
it is insufficient to make a strong conclusion that deviation absolutely vanishes in thermodynamic limit.
Nevertheless, the small discrepancy between $S^R$ and $S(\infty)$ is a correction to the finite size effect and thus it is irrelevant in the thermodynamic limit.

\begin{thebibliography}{99}
\bibitem{entropy}
In fact, one can rigorously prove that the energy fluctuation in random product states are comparable to that in random pure states
even though product states are a set of measure zero in the space of pure states.
\bibitem{longtime_ent}
We used two methods to estimate $S_L(\infty)$:  One is computing $S(t)$ at large $t$ such as 500, 1000, 10000 and seeing that $S(t)$
remains constant within the standard error at these long times.
The other is expanding the state in terms of the eigenstates of $H$ and giving each eigenstate a random phase (thus approximating infinite time)
and computing the entanglement entropy from the state with random phases.  We found these two methods give the same value within the error bar (fractional difference was less than 1.5$\%$).
\bibitem{longtime_diff}
We estimated $R(\infty)$ as the average of a few values of well-spaced $R(t)$ near and beyond the time scale set by the inverse many-body energy-level spacing.  For example, the inverse many-body level spacing for L = 14 is $\approx$ 533.  Then, $R(\infty)$ is estimated by averaging $R(450)$, $R(600)$, $R(750)$, $R(900)$, and $R(1050)$. Fractional fluctuation within that time range was as small as $0.12\%$ (this fractional fluctuation becomes smaller as we increase the system size).
\end{thebibliography}
\end{document}